\documentclass[doublecol]{epl2} 
\usepackage{graphicx}
\usepackage{color}
\usepackage{multirow}
\usepackage{hhline}
\usepackage[normalem]{ulem}

\title{Violation of Lee-Yang circle theorem for Ising phase transitions on complex networks}
\shorttitle{Violation of Lee-Yang theorem on complex networks} 

\author{M.~Krasnytska\inst{1,2} \and B. Berche\inst{2} \and Yu.~Holovatch\inst{1} \and R.~Kenna\inst{3}}
\shortauthor{M.~Krasnytska \etal}

\institute{
  \inst{1} Institute for Condensed Matter Physics of the National Acad. Sci. of Ukraine,
             79011 Lviv, Ukraine\\
  \inst{2} Institut Jean Lamour, CNRS/UMR 7198, Groupe de Physique
Statistique, Universit\'e de Lorraine, BP 70239, F-54506 Vand\oe
uvre-les-Nancy Cedex, France  \\
  \inst{2} Applied Mathematics Research Centre, Coventry University,
              Coventry, CV1 5FB, England
}

\abstract{The  Ising model on annealed complex networks with degree
distribution decaying algebraically as $p(K)\sim K^{-\lambda}$ has a
second-order phase transition  at finite temperature if $\lambda>
3$. In the absence of space dimensionality, $\lambda$ controls the
transition strength; mean-field theory applies for   $\lambda >5$
but critical exponents  are $\lambda$-dependent if $\lambda <  5$.
Here we show that, as for regular lattices, the celebrated Lee-Yang
circle theorem is obeyed for the former case. However, unlike on
regular lattices where it is independent of dimensionality,  the
circle theorem fails on complex networks when  $\lambda <  5$. We
discuss the importance of this result for  both theory and
experiments on phase transitions and critical phenomena. We also
investigate the finite-size scaling of Lee-Yang zeros in both
regimes as well as the multiplicative logarithmic corrections which
occur at $\lambda=5$.}
\begin{document}

\maketitle

Due to an abundance of natural and man-made network-like structures
\cite{networks}, phase transitions on complex networks are found in
systems ranging from  the physical   to the sociological and form a
fast developing theme of research \cite{Dorogovtsev08,sociophysics}.
Investigations have  centered on moments of the partition function
and how they scale close to critical points. Inspired by the
fundamental theorem of algebra, the Lee-Yang approach, on the other
hand, is based upon the zeros of the partition function, which it
interrogates   directly instead of through its moments
\cite{LeeYang52}. We investigate the Ising model on complex networks
through  Lee-Yang zeros, to reveal an unexpected feature of
importance for  theoretical and experimental studies.

For scale-free networks, the probability that a node has degree $K$
decreases for large $K$ as
\begin{equation}\label{2}
p(K)\sim K^{-\lambda}\, .
\end{equation}
Models on such networks have qualitatively similar critical behavior
to those on lattice systems. In particular, in the absence of a
space dimension $d$, the   exponent $\lambda$ determines collective
behaviour. There are lower and upper critical values of $\lambda$
for networks, analogous to critical dimensions for lattices. At the
upper critical values  scaling behaviour is modified by  logarithmic
corrections, while above them, critical exponents assume  mean-field
values \cite{Ginzburg}. (This analogy  has limitations; e.g., in
complex networks the mean number of second neighbors decreases with
increasing $\lambda$, while for  lattices it increases with
dimensionality.) Between the
 critical values,  critical exponents, amplitude ratios and
other universal quantities are $\lambda$- or $d$-dependent. For the
Ising model, as most models with discrete symmetry group, the lower
critical values are $\lambda_{\rm{lc}} = 3$, $d_{\rm{lc}}=1$, while
the upper ones are  $\lambda_{\rm{uc}} = 5$
 and $d_{\rm{uc}}=4$.

From its inception, the Lee-Yang approach  profoundly influenced
modern-day statistical mechanics \cite{Wu}. The  idea is to
generalize the magnetic field to the complex plane and  consider the
partition function as a polynomial in a related parameter
\cite{LeeYang52}. Fisher  extended this idea to the complex
temperature plane \cite{Fisherzeros}. The free energy is analytic in
regions free from zeros. This is important because phase transitions
are manifestations of nonanalyticities. These appear only when the
density of zeros accumulate in the thermodynamic limit. Moments such
as the susceptibility are directly expressible in terms of partition
function zeros which contain all of the thermodynamic information
accessible through standard approaches \cite{Bena}. In particular,
critical exponents are determinable through scaling of zeros near
the critical point and amplitude ratios from the angle at which they
impact onto the real parameter axis.

The full locus of partition function zeros contains additional
information. The  Lee-Yang theorem states that, under very general
conditions, all zeros are purely imaginary in the complex field
plane or on the unit circle after a change of variable. This  circle
theorem was proven for the Ising model on  lattices independent of
dimensionality $d$,  size, and boundary conditions of the lattice
and  range of the interactions. It has since been extended to a
wider class of regular-lattice  models. (For a review see
\cite{Bena}.) Recently, the Lee-Yang zeros of an Ising system were
experimentally accessed for the first time   using an approach which
relied crucially on the validity of the circle theorem \cite{Liu}.

The validity or otherwise of the circle theorem for lattice  systems
is dimensional independent. Given the parallel roles played by
$\lambda$ and $d$ outlined above, one may expect an analogous
statement to hold for annealed complex networks. However, here we
show that, while the theorem holds for the Ising model on a
 complex network provided $\lambda \ge \lambda_{\rm{uc}}$, it fails when  $\lambda < \lambda_{\rm{uc}}$.
This represents a striking difference between the $\lambda \ge
\lambda_{\rm{uc}}$ and
 $\lambda < \lambda_{\rm{uc}}$ fat-tailed cases and an unexpected
 difference between models on networks and lattices.
The finding has consequences for experimental realizations of
Lee-Yang zeros, and therefore for further investigations of phase
transitions, on networks. We discuss these consequences after
presenting our  analysis.

The Hamiltonian of the Ising model on a complex network is
\begin{equation}\label{1}
-{\cal H}=\frac{1}{2}\sum_{l\neq m}J_{lm}S_lS_m +H \, \sum _l S_l \,
.
\end{equation}
Here, $S_l=\pm 1$ is a spin variable, $H$  an external magnetic
field, the sum $\sum_{l\neq m}$ spans all pairs of $N$ nodes, and
the adjacency matrix $J_{lm}$ contains information about the network
structure in that  $J_{lm}=1$ if the nodes are linked and $J_{lm}=0$
otherwise. In general, the degree $K_l=\sum _m J_{lm}$ varies for
different $l$ and may be characterized by a   distribution $p(K)$.
For  annealed networks, the  links fluctuate on the same time scale
as the spin variables. Therefore the partition function is averaged
with respect to  the link distribution as well as the Boltzmann
distribution \cite{Brout59}. To implement this, each node $l$ is
assigned a hidden variable $k_l$, the  distribution of which is
given by  (\ref{2}).
The probability of a link between nodes $l$ and $m$ is then given by
$p_{lm}=k_lk_m/(N\langle k \rangle)$. It is easy to check that the
expected node degree  value is then ${\rm{E}} [K_l] = \sum_m p_{lm}
= k_l$. The above choice for $p_{lm}$ leads to the Hamiltonian
(\ref{1}) with separable interaction.

Applying the Hubbard-Stratonovich transformation,  averaging over
the spins can be performed exactly, leading to the representation
for the partition function,
\begin{eqnarray}\nonumber
Z_N (T,H) &=& \int_{-\infty}^{+\infty} \exp\Big \{\frac{-N\langle k
\rangle x^2}{2T}+ \\ \label{3} && \sum_{l}\ln \cosh[(
xk_l+H)/T]\Big\}dx \, ,
\end{eqnarray}
having dropped, for clarity, pre-factors insignificant for the
purpose of this study.

The sum over nodes $l$  may be rewritten in terms of a suitable
integral over $k$ with distribution function  $p(k)=c_\lambda
k^{-\lambda}$, in which $c_\lambda$ is a normalizing constant. One
obtains {{\cite{note}}}
\begin{equation}
Z_N (T,H)=\int_{0}^{+\infty} e^{-\frac{\langle k \rangle x^2T}{2}}
\Big ( e^{I^+_\lambda(x)
 } + e^{I^-_\lambda(x)} \Big) \,dx ,\label{4}
\end{equation}
where
\begin{equation}\label{5}
I^\pm_\lambda(x)= c_\lambda \frac{x^{\lambda-1}}
{N^{\frac{\lambda-3}{2}}}\, \int_{\frac{2x}{\sqrt{N}}}^{\infty}
 \frac{dy}{y^\lambda}\ln \cosh\Big(\pm y +\frac{H}{T} \Big) \, .
\end{equation}

This partition function sets the phase diagram of the Ising model on
an annealed scale-free network \cite{Dorogovtsev08,networks}. There
is no phase transition for $\lambda \le 3$. For $\lambda>3$ a
second-order  transition occurs at finite temperature $T_c=\langle
k^2 \rangle/\langle k \rangle$. In the region $3<\lambda< 5$ it is
governed by  $\lambda$-dependent exponents which, in  standard
notation for  the thermal behaviour of the specific heat,
magnetization, susceptibility and the field-dependency of the
magnetization, are \cite{Dorogovtsev02}
\begin{equation}\label{6}
\alpha=\frac{\lambda-5}{\lambda-3} , \hspace{0.5em}
\beta=\frac{1}{\lambda-3}, \hspace{0.5em}  \gamma=1, \hspace{0.5em}
\delta=\lambda-2\, .
\end{equation}
For $\lambda>5$ the exponents attain their usual Ising mean-field
values $\alpha = 0$, $\beta = 1/2$, $\gamma = 1$, and $\delta = 1/3$
enhanced by the logarithmic corrections at $\lambda=5$. Thus the
global variable $\lambda$ determines  the critical behaviour in a
manner similar to the space dimensionality for lattice models.

We are interested in the zeros of the partition function (\ref{4})
at the critical temperature. Substitute $T=T_c$ in (\ref{4}) and
make an asymptotic expansion keeping the leading terms in $1/N$ and
$H$, to find
\begin{equation}\label{7}
 Z(h)=\left\{
\begin{array}{lll}
                 \int_{0}^{+\infty}{ e^{-x^{\lambda-1}}     \cosh(hx)dx}\, , && 3<\lambda<5,
                \\
                 \int_{0}^{+\infty}{ e^{-x^4}      \cosh(h x)dx } \, , && \lambda\geq 5\, .
              \end{array}
  \right.
\end{equation}
Here, the $H$- and $N$-dependencies  are adsorbed in a single
variable $h$, the explicit form for which differs in different
regions of $\lambda$:
\begin{eqnarray}\label{8}
 h=\left\{
\begin{array}{ccc}
                & H \frac{\langle k \rangle^2}{\langle k^2 \rangle}a(\lambda)^{1/(1-\lambda)}N^{\frac{\lambda-2}{\lambda-1}}\, , & 3<\lambda<5, \\
                & H \frac{\langle k \rangle^2}{\langle k^2 \rangle}\Big(\frac{24}{\ln N}\Big)^{1/4}N^{3/4}\ \, , & \lambda=5, \\
                & H \frac{\langle k \rangle^2}{\langle k^2 \rangle}\Big(\frac{12}{\langle k^4 \rangle}\Big)^{1/4}N^{3/4}\, , & \lambda>5\, .
              \end{array}
  \right.
\end{eqnarray}
Note that $Z(h)$ is independent of $\lambda$ when $\lambda \geq 5$.
In particular, the logarithmic corrections at $\lambda = 5$ reside
in the form of $h$ in the middle expression of (\ref{8}). Therefore,
working in terms of the reduced variable $h$, the loci of partition
function zeros for $\lambda = 5$ and $\lambda>5$ are the same.
 For $\lambda <5$ the values of
the coefficients $a(\lambda)=-c_\lambda\int_0^\infty dy\,
y^{-\lambda}(\ln\cosh y -y^2/2) >0$ can be found numerically and are
listed  in \cite{Krasnytska13}.

\begin{figure}[t]
\centerline{\includegraphics[angle=0,
width=4.4cm]{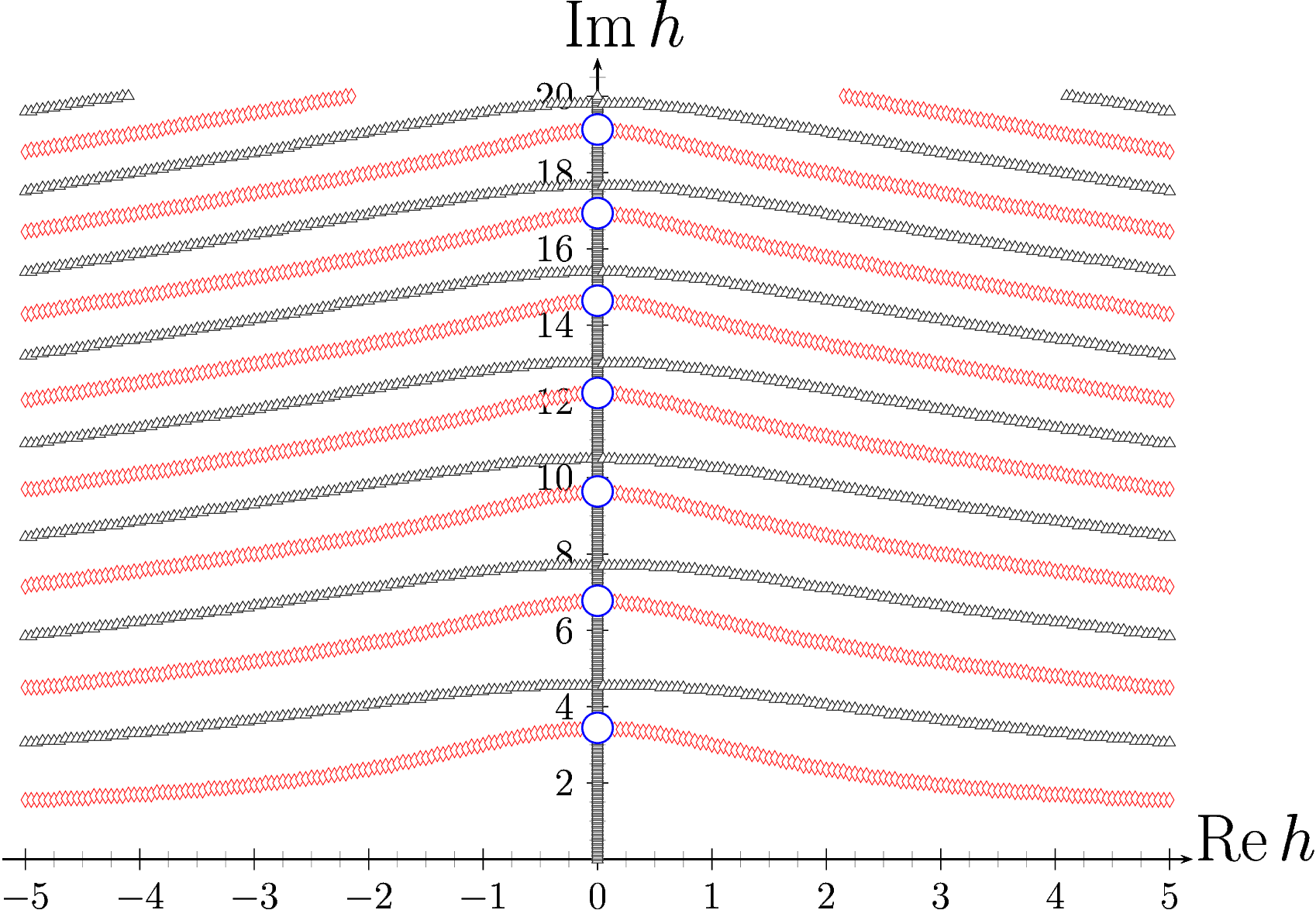}~~~\includegraphics[angle=0,
width=4.4cm]{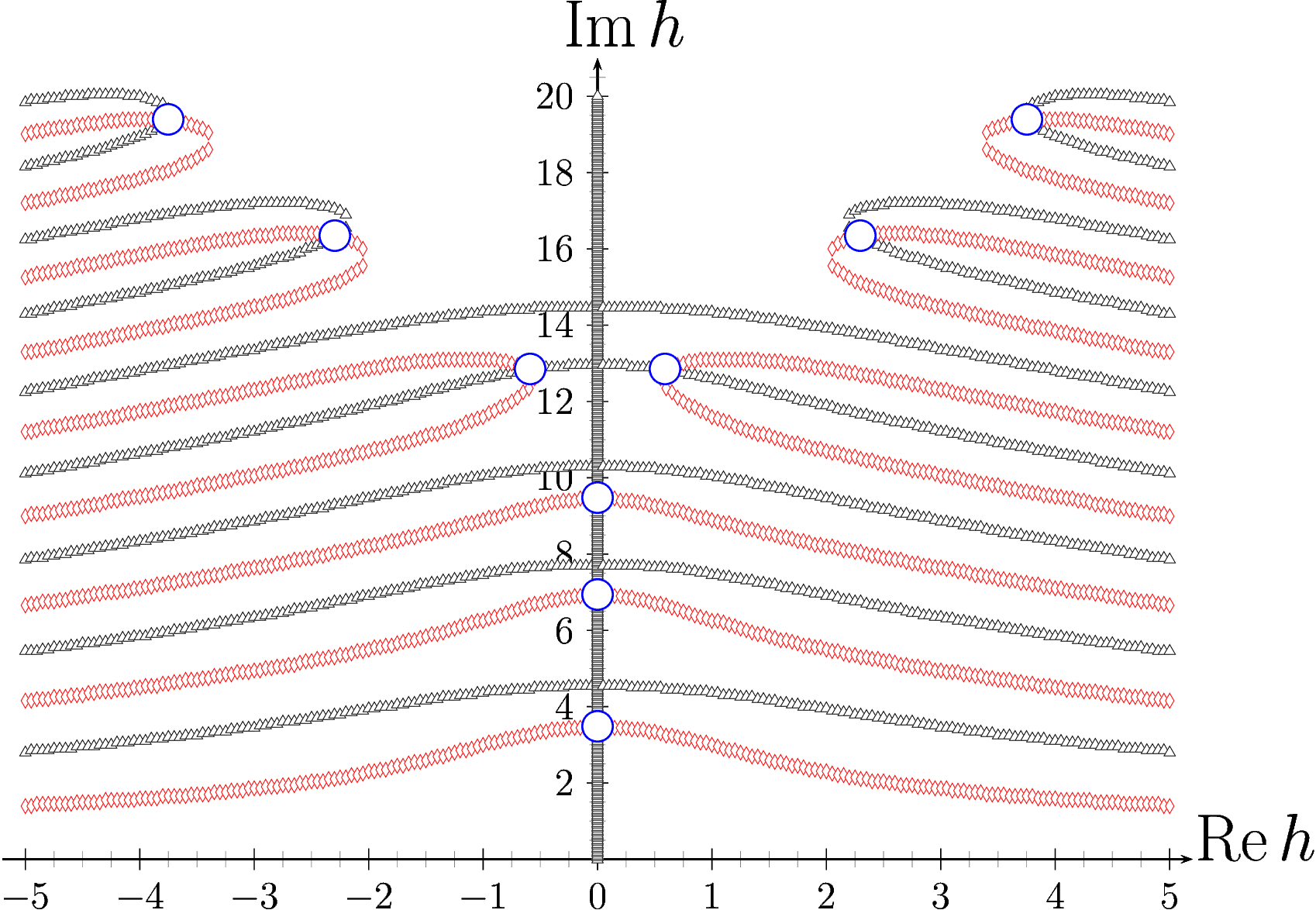}}
 \vspace*{1pt}
\caption{Solutions of the system of equations ${\rm{Re}}~Z(h)=0$
(diamonds, red online) and ${\rm{Im}}~ Z(h)=0$ (triangles, including
along the imaginary axis) for the partition function (\ref{7}) when
$\lambda{ {\geq}}5$ (left) and $\lambda=4.5$ (right). The Lee-Yang
zeros are given by the intersections of contours of different types
and are denoted by circles  (blue online). All zeros are imaginary
only when $\lambda\geq 5$.
 \label{fig1}}
\end{figure}

Let us now consider the  partition function (\ref{7}) in  complex
magnetic field $h={\rm{Re}}~h + i~{\rm{Im}}~ h $. We are  interested
in its zeros $h_j$ at the critical point. The index $j$ identifies
the sequence of zeros relative to the origin in the upper half-plane
with $j=1$ identifying zero closest to $h=0$. Increasing  $N$, the
zeros approach the origin and their  scaling at criticality is
governed by the exponent $\sigma$ \cite{Itzykson83}:
\begin{equation}\label{9}
H_j(N)\sim\Big ( \frac{j}{N} \Big )^{\sigma} \, .
\end{equation}
Comparing  (\ref{8}) and (\ref{9})  one finds
\begin{eqnarray}\label{10}
 \sigma=\left\{
\begin{array}{ccc}
                & \frac{\lambda-2}{\lambda-1}\, , & 3<\lambda \le 5, \\
                 & 3/4 \, , & \lambda \ge 5\, .
              \end{array}
  \right.
\end{eqnarray}

Therefore, $\sigma$ becomes $\lambda$-dependent in the region
$3<\lambda<5$, similar to the other critical exponents (\ref{6}).
Moreover,  the validity of the scaling relation
$\sigma=\beta\delta/(2-\alpha)$ is verified \cite{Itzykson83}. Also,
from (\ref{8}), a logarithmic correction appears in the marginal
case $\lambda=5$: $H_j \sim N^{-3/4} (\ln N)^{1/4}$. Again, the
power of the logarithm complies with the corresponding scaling
relation \cite{logcorrections}.

\begin{largetable}
\caption{The  first five  zeros $h_j$ for different values of
$\lambda$. \label{tab1}}
\begin{center}
\begin{tabular}{|c|c|c|c|c|}
  \hline
        &  $\lambda{ \geq}5$ & $\lambda=4.5$ &  $\lambda=4$    &  $\lambda=3.5$ \\
  \hline
  $j=1$& $i3.495$ & $i3.495$  & $i3.569$ &  $i3.762$  \\
  \hline
 $j=2$  &$i6.784$ & $i6.933$  &  $i7.823$ & $1.875+i7.212$   \\
  \hline
 $j=3$ & $i9.636$ &  $i9.474$ & $2.418+i11.466$  & $3.659+i9.496$      \\
  \hline
  $j=4$  & $i12.229$ &  $0.589+i12.848$   & $4.014+i14.174$  &  $5.138+i11.351$      \\
\hline
  $j=5$  & $i14.650$ &  $2.297+i16.346$  & $5.446+i16.574$   &  $6.435+i12.983$      \\
  \hline
\end{tabular}
\end{center}
\end{largetable}

The contours in Fig.~\ref{fig1} depict curves in the complex plane
along which the real and imaginary parts of the partition functions
(\ref{7}) vanish at $T=T_c$ for $\lambda \geq 5$ and $3 < \lambda <
5$, exemplified by  $\lambda = 4.5$.
 Note that  ${\rm{Im}}Z(h)$ vanishes
when $h$ itself is imaginary because the partition function is an
even function of $h$.
   The intersections of the different contours give the locations
of the Lee-Yang zeros. When $\lambda \ge 5$  all zeros plotted are
on the imaginary axis. But for $\lambda=4.5$ only the first three
zeros are imaginary and zeros of higher order ($h_j$ for $j>3$) have
non-vanishing real parts. This type of behaviour is found for all
other values of $\lambda$ in the region $3<\lambda<5$; there is a
finite number, $\mathcal{N}$, such that $h_j$ is imaginary for $j\le
{\mathcal{N}}$ and has non-vanishing real part for $j>\mathcal{N}$.
Moreover, $\mathcal{N}$ decreases with decreasing $\lambda$. The
numerically determined values of Lee-Yang zeros are listed in Table
\ref{tab1} for various values of $\lambda < \lambda_{\rm{uc}} = 5$,
and for $\lambda { \geq} 5$.  When $\lambda=4$, $\mathcal{N}=2$
while for $\lambda = 3.5$, $\mathcal{N}=1$.

\begin{figure}[t]
\centerline{\includegraphics[angle=0, width=5cm]{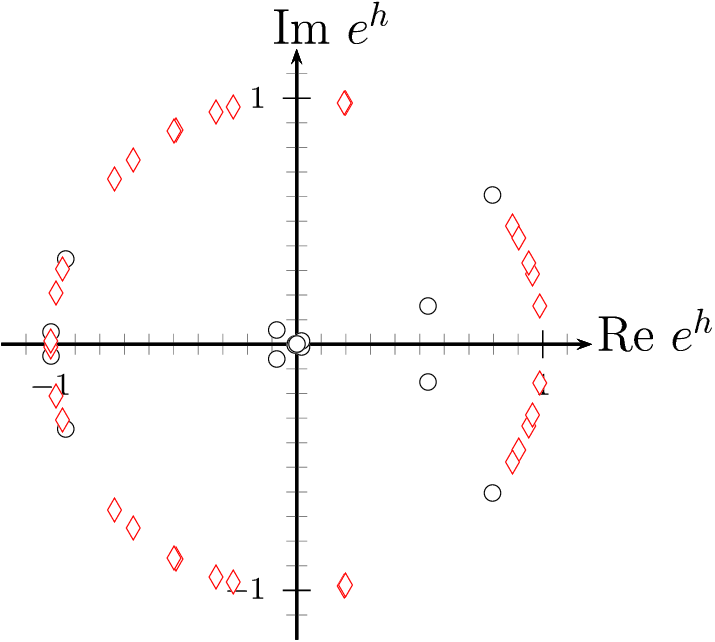}}
 \vspace*{1pt}
\caption{ Lee-Yang zeros  in the complex $e^h$-plane. The diamonds
(red online) correspond to $\lambda{ \geq}5$ and the circles to
$\lambda=4.5$. The plot depicts only those zeros that fit into the
region shown. \label{fig2}}
\end{figure}

The picture is reaffirmed in Fig.~\ref{fig2} which demonstrates that
the  circle theorem holds in the complex $e^h$-plane for the Ising
model on the annealed scale-free network for  $\lambda \geq 5$.
However it does not hold for $\lambda <  5$, where only a few low
index zeros lie on the unit circle, the rest being scattered on the
complex plane.

Thus, while our numerical examples hint that the Lee-Yang theorem
holds for $\lambda \geq 5$, we have established  the surprising
result that it does not hold when $\lambda <5$. To gain analytical
insight, we can determine an asymptotic form for the partition
function (\ref{7}) when ${\rm{Re}}~h=0$
in the limit of large ${\rm{Im}}~h$. We employ the Erdelyi lemma;
if
\begin{equation}\label{11}
F(y)= \int_0^a x^{\beta-1}f(x)e^{iyx^\alpha}dx,
\end{equation}
where $\alpha\geq1$, $\beta>0$ and the function $f(x)$  together
with all  derivatives vanishes at the upper integration limit, the
large-$y$ asymptotic behaviour  is \cite{Fedoruk87}
\begin{equation}\label{12}
F(y)\sim \sum_{k=0}^\infty a_k y^{-\frac{k+\beta}{\alpha}}\, ,
\end{equation}
where
$a_k=\frac{f^{k}(0)}{k!\alpha}\Gamma\left({\frac{k+\beta}{\alpha}}\right)
\exp\left[{\frac{i\pi(k+\beta)}{2\alpha}}\right]$.
 Putting $h=i r$ 
one can represent the partition function for $3<\lambda<5$ in the
form  (\ref{11}):
\begin{equation}\label{13}
Z(ir) = \frac{5}{\lambda-1}{\rm{Re}}~\int_0^\infty
x^{\frac{5}{{\lambda-1}}-1}e^{-x^5}e^{ir x^{5/({\lambda-1})}}dx\, .
\end{equation}
From (\ref{12}), the asymptotic expansion is
\begin{equation}\label{14}
Z(ir)\sim (\lambda-1)\sum_{k=0}^\infty b_k r^{-\frac{k
(\lambda-1)}{5}-1}\, ,\hspace{0.5cm} r\rightarrow\infty \,,
\end{equation}
with coefficients
\begin{equation}\label{15}
b_k=\frac{f^{k}(0)}{5k!}\Gamma\left[{\frac{k(\lambda-1)}{5}+1}\right]
\cos\left[{\frac{\pi(k(\lambda-1)+5)}{10}}\right]\, .
\end{equation}

 The  asymptotics for the $\lambda\geq5$
partition function (\ref{7})  can be evaluated by steepest descent,
leading to
\begin{eqnarray}\nonumber
Z(ir) &\sim & \exp{
     \left[{
                -\frac{3}{2}
                         \left({\frac{r}{4}}\right)^{\frac{4}{3}}
         }\right]
        }
\cos{
     \left[{
               \frac{3\sqrt{3}}{2}
                     \left({\frac{r}{4}}\right)^{\frac{4}{3}}
         }\right]
    } \, , \\ \label{17} && \hspace{0.5cm} r \rightarrow\infty \, , \hspace{0.5em} \lambda\geq5 .
\end{eqnarray}
The  trigonometric function in (\ref{17}) ensures the partition
function alters in sign as a function of $r$. This signals that it
vanishes on the imaginary axis for large (infinite) values of $r$,
suggesting that $\mathcal{N}$ is infinite. If the pattern of zeros
observed in Fig.\ref{fig1} is generic  in that $h_j$ has a
non-vanishing real part only when $j > \mathcal{N}$, this indicates
that all zeros are on the imaginary-$h$ axis, suggesting
 that the Lee-Yang circle theorem is obeyed in the complex fugacity plane.

\begin{figure}[t]
\centerline{\includegraphics[angle=0, width=5cm]{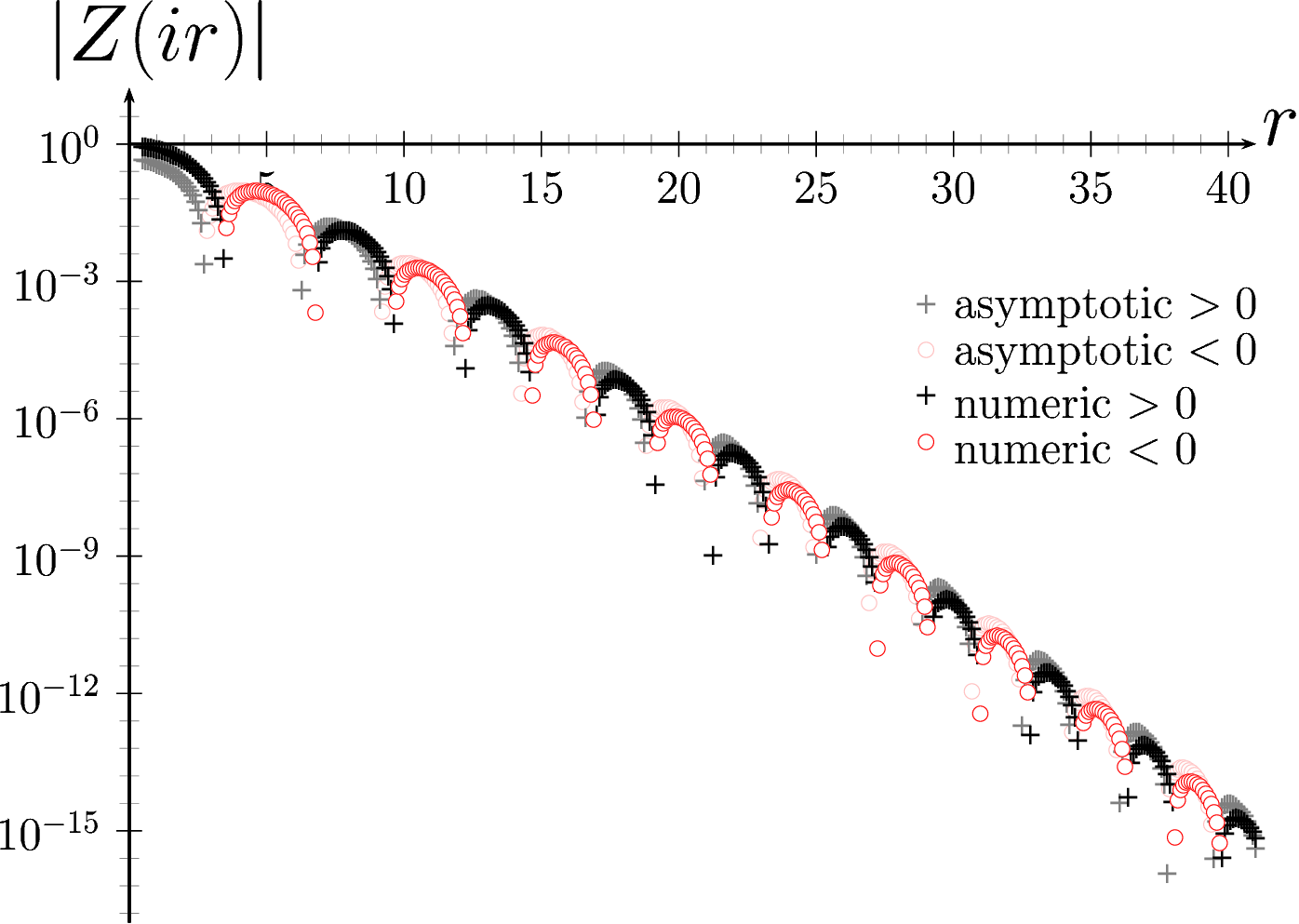}}
 \centerline{{\bf a}}
 \centerline{\includegraphics[angle=0, width=5cm]{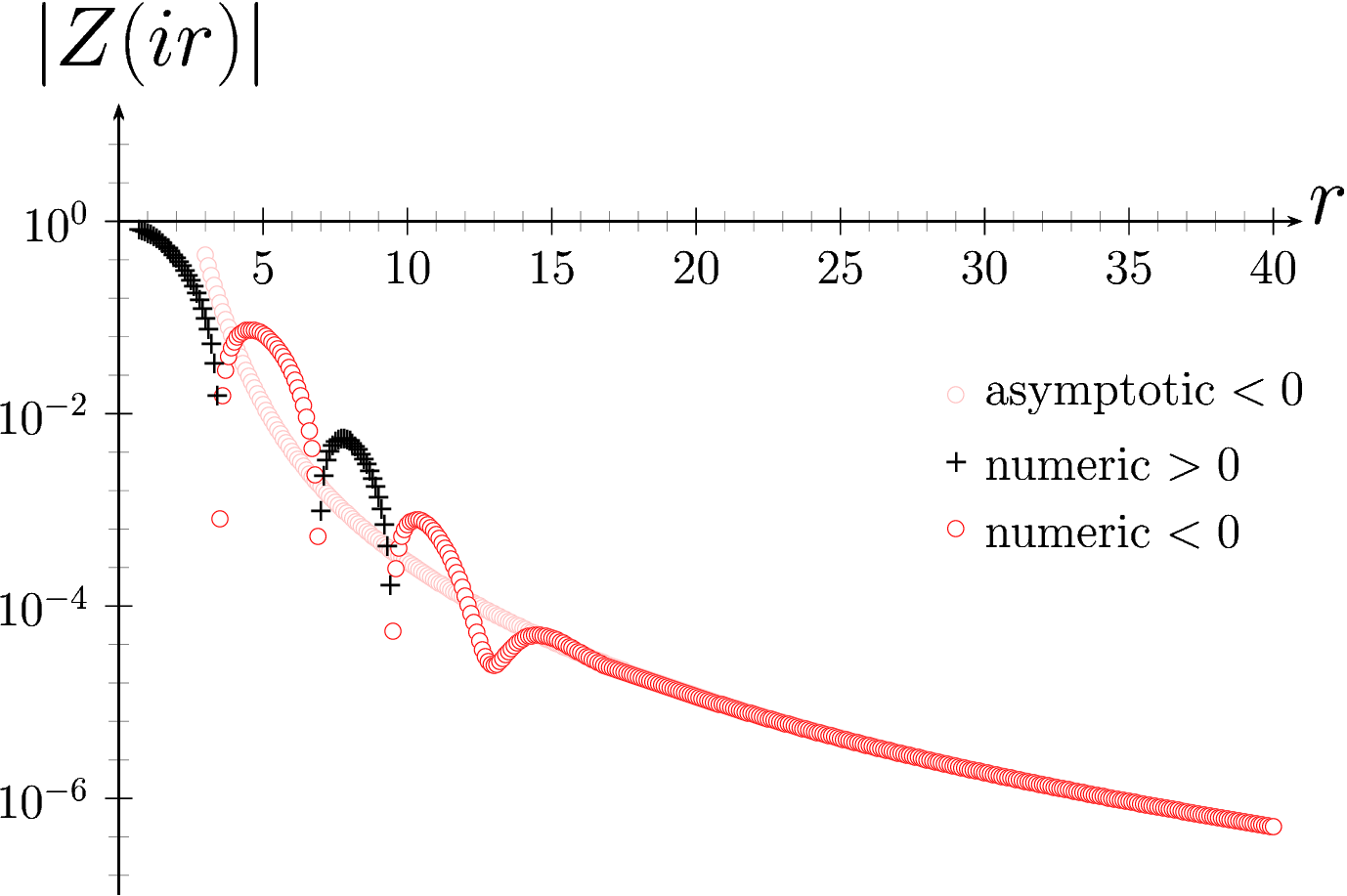}}
 \centerline{{\bf b}}
 \vspace*{1pt}
\caption{Numerically calculated absolute values of the partition
function along the imaginary axis: $|Z (ir)|$ with $r={\rm{Im}}\,h$.
The asymptotic expansions using Eq. (\ref{14}) and (\ref{17})  in
the limit $r \to \infty$ are also plotted. The symbols ``+''
 indicate $Z (ir)>0$ (black and gray online correspondingly)  and circles
 indicate $Z (ir)<0$ (red and pink red online).  Panel (a) is for $\lambda\geq5$ and Panel
(b) for $\lambda=4.5$. The number of zeros is finite for
$\lambda<5$. \label{fig3}}
\end{figure}

In Fig. \ref{fig3} we compare numerically calculated absolute values
  of the integral $|Z(ir)|$ (\ref{7})  with its asymptotic expansions
    (\ref{14})  and (\ref{17}) for different $\lambda$. The behaviour for
$\lambda \ge 5$ is qualitatively different from that at $\lambda
<5$: whereas in the former case the function oscillates even in the
asymptotic regime (meaning that  the number of zeros is unbounded),
this is not the case for $\lambda <5$. There, after a finite number
of oscillations the function approaches its asymptotic, bounding the
regime of zeros. Moreover, the number of oscillation decreases with
decreasing $\lambda$.

In summary, besides delivering  finite-size scaling, we found an
unexpected result for the locus of Lee-Yang zeros for the Ising
model on annealed scale-free networks,   namely, while   the circle
theorem  is obeyed for $\lambda \ge \lambda_{\rm{uc}} = 5$, it fails
for $\lambda < \lambda_{\rm{uc}} = 5$. This is surprising because it
means that the similarity between  roles played by  $\lambda$ (for
complex networks) and $d$ (for regular lattices), which control
universal aspects of phase transitions, do not extend to the
fundamental Lee-Yang level.

Recently Peng et al. related imaginary magnetic fields associated
with a  bath of Ising spins to the quantum coherence of a probe
\cite{Liu}. This allowed them to  identify   the experimentally
obtained times at which the quantum coherence disappears as
imaginary  Lee-Yang zeros. This demonstration of a physical
manifestation of Lee-Yang zeros is important at a fundamental level
and points to new, experimental ways of studying zeros in
many-bodied materials. However, an identification of the type made
in \cite{Liu} relies crucially on the Lee-Yang theorem in that it
can only be made for  imaginary zeros \cite{WeiLiu}. The results
herein show that while these are accessible for lattices and complex
networks for sufficiently large $\lambda$, not all network zeros are
accessible in this manner for $\lambda < 5$. A challenge for
experiment is to find another way to access them.

Finally we address  the question of why the theorem is obeyed for
some but not all values of $\lambda$. It was proven \cite{Lieb81}
that the theorem holds  for any Ising-like model with ferromagnetic
interactions. One may therefore anticipate that the Hamiltonian
(\ref{1}) for a single network realization should have the Lee-Yang
property, since the adjacency matrix elements $J_{ij}$ are
non-negative. However, averaging over an ensemble of networks
amounts to  taking  a sum (or  integral) of functions and there is
no guarantee for $\lambda<5$ that the zeros of the sum preserves the
property. Indeed, one can consider heuristically the Landau  free
energy for scale-free networks, $F(M) \sim M^2 + M^{\lambda-1} + M^4
+ \dots $ \cite{Goltsev03}. When $\lambda < 5$ the second term on
the right is dominant for small order parameter $M$, delivering
$\lambda$-dependent critical exponents. The partition function is
also $\lambda$-dependent and its zeros are complex. When $\lambda >
5$, however,  the first  term on the right  dominates. This leads to
mean-field critical behaviour which is independent of $\lambda$.
Since mean-field theory obeys the Lee-Yang theorem, so too must
averaged annealed complex networks with $\lambda > 5$.

\section{Acknowledgements}
This work was supported in part by FP7 EU IRSES projects No.$269139$
(DCP-PhysBio), No. $295302$ (Spider),
         IIF Project  No.$300206$ (RAVEN),
                and by the Coll\`ege Doctoral ``Statistical Physics of Complex Systems''
Coventry-Leipzig-Lviv-Nancy. We thank Yuri  Kozitsky, Taras
Krokhmalskii,  Jean-Yves Fortin and Nikolay Izmailian  for comments
and discussions.


\begin{thebibliography}{99}

\bibitem{networks}
Newman M., Networks: An Introduction (Oxford University Press) 2010.


\bibitem{Dorogovtsev08}
Dorogovtsev S.~N., Goltsev A.~V., and Mendes J. F. F., \textsl{Rev.
Mod. Phys.}, \textbf{ 80} (2008) 1275.


\bibitem{sociophysics}
Galam S. Sociophysics: A Physicist's Modeling of Psycho-Political
Phenomena (Understanding Complex Systems) 2012.



\bibitem{LeeYang52} \
Yang C. N. and  Lee T. D., \textsl{Phys. Rev. Lett.} \textbf{ 87},
(1952) 404; Lee T. D. and Yang C. N., ibid., (1952) 410.



\bibitem{Ginzburg}
Ginzburg V. L., \textsl{Fiz. Twerd. Tela}, \textbf{2} (1960)  2031
[\textsl{Sov. Phys. Solid State}, \textsl{2} (1961) 1824].

\bibitem{Wu}
Wu F.Y.,
\textsl{Int. J. Mod.  Phys.~B}, \textbf{ 22} (2008)  1899.

\bibitem{Fisherzeros}
 Fisher M.E., in: Lectures in Theoretical Physics, ed. by W. E. Brittin (New York: Gordon and
 Breach), vol. VIIC, 1968.

\bibitem{Bena}
 Bena I., Droz M. and Lipowski A., \textsl{Int. J. Mod.  Phys.~B}, \textbf{19}
(2005) 4269.


\bibitem{Liu}
Peng X., Zhou H., Wei Bo-Bo, Cui J., Du J., and Liu R.-B.,
 \textsl{Phys. Rev. Lett.}, \textbf{ 114} (2015) 010601.


\bibitem{Brout59}
Brout R., \textsl{Phys. Rev.}, \textbf{ 115} (1959) 824.

\bibitem{note}
We keep  contributions in the integral representation (\ref{5})
which are leading in $N$. Only the higher-order terms  depend on the
form of the $N$-dependency of the upper cut-off $k_{\rm max}$. Also,
as usual, we choose the lower cut-off value $k_{\rm min}=2$ to keep
the network connected.

\bibitem{Dorogovtsev02}
Dorogovtsev S. N., Goltsev A. V., and Mendes J. F. F., \textsl{Phys.
Rev. E}, \textbf{66} (2002) 016104; Leone M., V\'azquez A.,
Vespignani A., and Zecchina R.,  \textsl{Eur. Phys. J. B}, \textbf{
28} (2002) 191.

\bibitem{Krasnytska13}
Krasnytska M., Berche B., and Holovatch Yu., \textsl{Condens. Matter
Phys.}, \textbf{ 16} (2013) 23602.

\bibitem{Itzykson83}
Itzykson C., Pearson R. B., and Zuber J. B., \textsl{Nucl. Phys. B},
\textbf{220} [FS8] (1983) 415.

\bibitem{logcorrections}
Kenna R., in: Order, Disorder, and Criticality, ed. by Yu. Holovatch
(World Scientific, Singapore), vol. 3, 2013, pp. 1-46.

\bibitem{Fedoruk87}
Fedoryuk M. V.,  Asymptotic Methods in Analysis. Analysis I
Encyclopaedia of Mathematical Sciences, ed. by R. V. Gamkrelidze
(Springer Berlin Heidelberg), vol. 13, 1989, pp. 83-191.


\bibitem{WeiLiu} Wei B.-B.~and Liu R.-B.,
\textsl{Phys. Rev. Lett.}, \textbf{109} (2012) 185701.


\bibitem{Lieb81}
Lieb E. H. and Sokal A. D., \textsl{Commun. Math. Phys.}, \textbf{
80} (1981) 153.

\bibitem{Goltsev03}
Goltsev A. V., Dorogovtsev S.N., and Mendes J. F. F., \textsl{Phys.
Rev. E }, \textbf{ 67} (2003) 026123.
 \end{thebibliography}
\end{document}